\documentclass[conference]{IEEEtran}
\IEEEoverridecommandlockouts

\usepackage[T1]{fontenc}
\usepackage[utf8]{inputenc}
\usepackage{lmodern}
\usepackage{amsmath}
\usepackage{graphicx}
\usepackage{booktabs}
\usepackage{tabularx}
\usepackage{array}
\usepackage{xcolor}
\usepackage[hyphens]{url}
\usepackage[hidelinks]{hyperref}
\usepackage{cite}
\usepackage{tikz}
\usetikzlibrary{arrows.meta,fit,backgrounds}

\newcommand{\code}[1]{\path{#1}}
\newcolumntype{Y}{>{\raggedright\arraybackslash}X}

\begin{document}

\title{Optional Traps: An Empirical Study of LLM Agents Against Credential Deception and Alert-Driven Revocation}

\author{\IEEEauthorblockN{Raditio Ghifiardi}
\IEEEauthorblockA{Independent Researcher\\
Jakarta, Indonesia}}
% TODO before submission: add a contact e-mail if desired.

\maketitle

\begin{abstract}
Deception (honeypots, decoys and canary credentials) rests on a simple premise: legitimate users never touch a fake asset, so any touch is a high-confidence signal. We ask whether that premise holds when the attacker is a large language model (LLM) agent, and what deception actually contributes once ordinary hardening is in place. We built an isolated, hash-chained lab around a mock Records API and report four studies. (1) In an exploratory run, a locally hosted 9.7B-parameter model ignored every optional decoy, read a genuinely leaked service key from a debug endpoint and exported the protected records in both static and adaptive deception modes. (2) After redesigning the defense to remove the leak, plant a canary credential at the former leak and revoke the offending client, a locally frozen 60-session study with a 4B-parameter model found genuine export success of 20/20 at baseline, 0/20 with hardening alone and 0/20 with hardening plus deception (exact paired McNemar $p = 1.9\times10^{-6}$ for baseline versus defended; $p = 1$ for hardened versus defended). Deception added detection and per-client revocation in 19 of 20 defended sessions, but no measurable prevention beyond hardening. (3) On Google Cloud, a real OpenCanary login alert was authenticated, signed and turned into revocation of the attacking client; the first denied export followed within an observed upper bound of 303\,ms, while a separate legitimate client completed 100/100 requests, and all 14 control checks passed. (4) In multi-agent swarm engagements, agents that used the canary were revoked 39--59\,s after the bait was served without affecting siblings, but one model-driven legitimate agent also took the bait, illustrating a false-positive risk specific to LLM clients. We conclude that against LLM agents, traps must sit on the attacker's path rather than beside it, that closing real weaknesses does the preventive work, and that deception's measurable value is fast, attributable detection. All results are lab-scale and are not a production claim.
\end{abstract}


\section{Introduction}
LLM agents can now chain reconnaissance, credential discovery and exploitation against web targets with little human guidance~\cite{fang2024hack}. Defenders have long relied on deception to detect such activity early: honeypots~\cite{spitzner2002,stoll1989}, honeytokens and honeywords~\cite{juels2013honeywords}, and planned adversary engagement~\cite{mitre-engage}. Recent work argues that LLM agents may be especially susceptible to deception because they read and act on whatever text the environment returns~\cite{ayzenshteyn2025cheat,pasquini2024mantis,reworr2024honeypot}.

This paper tests that intuition in a controlled lab and asks what deception contributes when it is measured against an equivalent defense without it. We pose three research questions:
\begin{itemize}
\item \textbf{RQ1 (susceptibility).} Does an LLM agent take optional decoys and canary credentials when a genuine weakness is also reachable?
\item \textbf{RQ2 (marginal value).} Given ordinary hardening, does adding deception further reduce an LLM agent's success at exfiltration?
\item \textbf{RQ3 (response).} Can a real honeypot alert be turned into fast, attributable and narrowly scoped enforcement without harming legitimate clients?
\end{itemize}

\textbf{Contributions.}
\begin{enumerate}
\item A reproducible, isolated deception lab with hash-chained event logs, signed control events and per-client revocation, evaluated both locally and on Google Cloud.
\item A documented bypass in which a live model ignored all optional decoys and used a genuine leak (Study~1), and the defense redesign it motivated.
\item A locally frozen, paired 60-session study with an ablation that separates the effect of hardening from the effect of deception (Study~2). Hardening alone accounted for all observed prevention; deception's measured contribution was detection and revocation.
\item A 14-check cloud validation of alert-driven revocation using a real OpenCanary sensor (Study~3), and exploratory multi-agent swarm engagements that surfaced a false-positive mode in which a legitimate LLM client takes the bait (Study~4).
\end{enumerate}

We report negative and null results as they occurred. Small samples, a single target and local models of 4B and 9.7B parameters bound every claim; Section~\ref{sec:limits} lists these limits.

\section{Background and Related Work}
Prior work establishes that deception detects human intruders and argues it should also work against LLM agents; few studies measure it against a no-deception control with the same hardening.

\textbf{Deception for human attackers.} Honeypots expose systems with no production purpose, so any interaction is suspect~\cite{spitzner2002,stoll1989}. Honeytokens extend the idea to data: honeywords plant decoy passwords so that a cracked password file reveals itself when a decoy is used~\cite{juels2013honeywords}. MITRE Engage frames denial and deception as planned adversary engagement with defined goals~\cite{mitre-engage}. OpenCanary is an open-source honeypot daemon that emits alerts on interaction~\cite{opencanary}; we use it as the real sensor in Studies~1, 3 and~4.

\textbf{LLM agents as attackers.} Fang et~al.\ showed that tool-using LLM agents can autonomously exploit vulnerable web applications, including multi-step attacks, without prior knowledge of the specific flaw~\cite{fang2024hack}. This motivates defenses that assume the attacker reads and reasons over server responses.

\textbf{Deception aimed at LLM agents.} Pasquini et~al.\ propose Mantis, which embeds prompt-injection payloads in decoy services to derail or counter-attack LLM-driven attackers~\cite{pasquini2024mantis}. Ayzenshteyn et~al.\ present CHeaT, proactive defenses that cloak assets, plant honey and set traps tailored to LLM penetration-testing agents~\cite{ayzenshteyn2025cheat}. Reworr and Volkov deploy an internet-facing honeypot with embedded prompt injections and timing analysis to detect AI hacking agents in the wild~\cite{reworr2024honeypot}.

\textbf{How this work differs.} We deliberately avoid prompt injection and hidden instructions: our decoys are ordinary fake assets and a canary credential, which keeps the defense independent of model-specific prompt behaviour. Our main methodological addition is the hardened-versus-defended ablation in Study~2, which separates prevention due to removing a real weakness from prevention due to deception. We also tie the honeypot alert to an authenticated, client-scoped enforcement action and measure its effect on a co-resident legitimate client.

\section{Lab Design and Threat Model}
\label{sec:design}
The lab exposes one mock Records API with a deliberate weakness and measures every outcome from the target's own ground truth.

\textbf{Target.} A Records API serves synthetic records. Its genuine objective is \code{GET /api/v2/records/export} with a valid bearer key. The deliberate weakness is a debug endpoint (\code{/api/v1/debug/env}) that, in V1 and in the V2 baseline, leaks the real service key. All addresses are documentation ranges or internal Docker addresses, and all credentials are synthetic.

\textbf{V1 deception.} Static mode adds a backup file (\code{/backup/config.bak}) holding a canary key and a deprecated internal API listed in \code{robots.txt} and the OpenAPI document. Adaptive mode makes the internal API return endless pagination leads. Use of the canary key, or 12 decoy hits, terminates the session inside the API. These decoys sit \emph{beside} the genuine path.

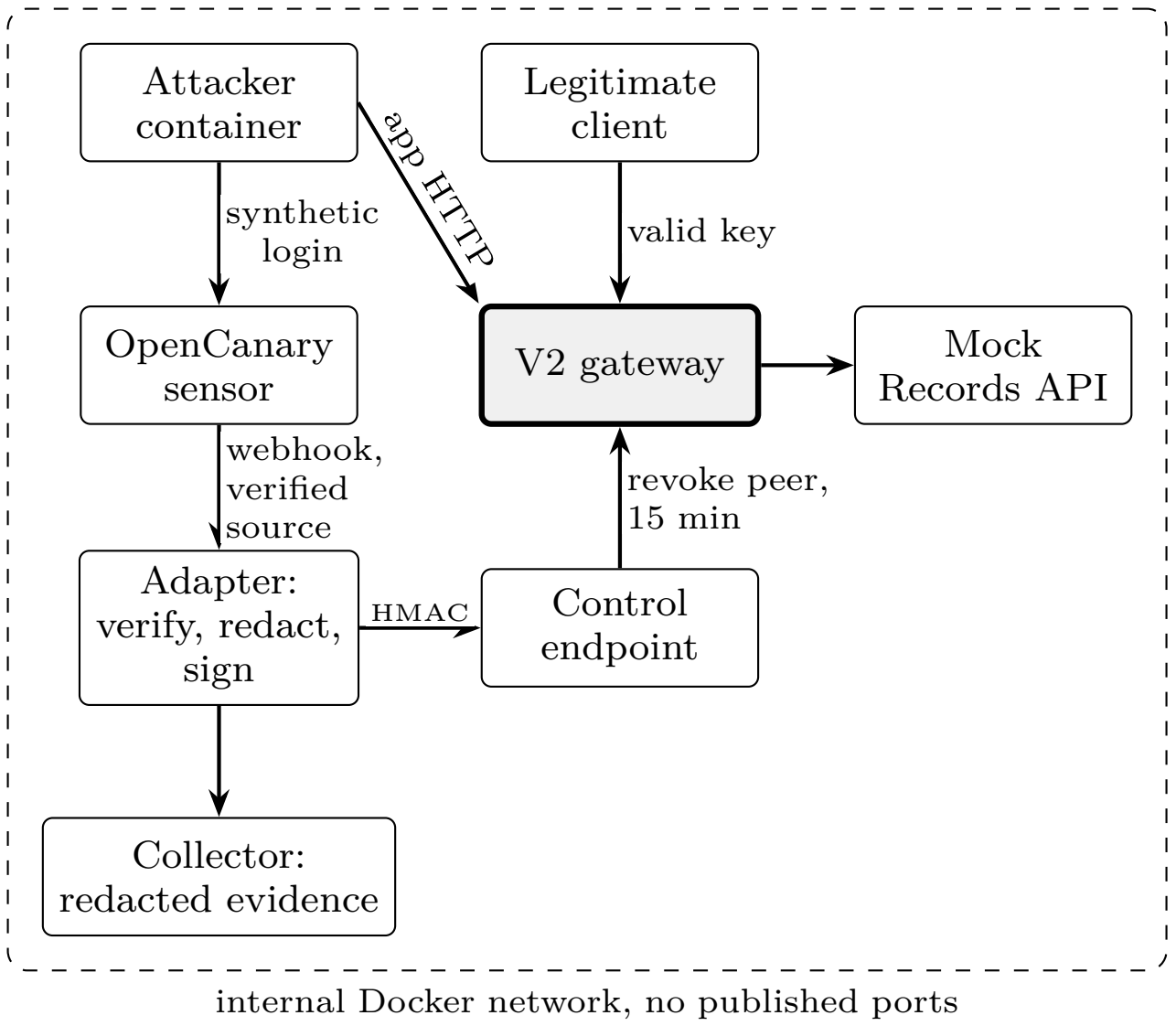
\begin{figure}[t]
\centering
% Figure 1: V2 architecture and trust boundaries (TikZ, single column).
\begin{tikzpicture}[
  font=\footnotesize,
  box/.style={draw, rounded corners=2pt, align=center, minimum width=2.0cm, minimum height=0.85cm, fill=white},
  key/.style={box, very thick, fill=gray!12},
  arr/.style={-{Stealth[length=2mm]}, thick},
  lbl/.style={font=\scriptsize, fill=white, inner sep=1pt, align=center}
]
  \node[box] (atk) at (0,0) {Attacker\\container};
  \node[box] (leg) at (2.9,0) {Legitimate\\client};
  \node[box] (oc)  at (0,-1.9) {OpenCanary\\sensor};
  \node[key] (gw)  at (2.9,-1.9) {V2 gateway};
  \node[box] (api) at (5.6,-1.9) {Mock\\Records API};
  \node[box] (ad)  at (0,-3.8) {Adapter:\\verify, redact,\\sign};
  \node[box] (ctl) at (2.9,-3.8) {Control\\endpoint};
  \node[box] (col) at (0,-5.6) {Collector:\\redacted evidence};

  \draw[arr] (atk) -- node[lbl, right] {synthetic\\login} (oc);
  \draw[arr] (atk.east) -- node[lbl, sloped, above] {app HTTP} (gw.north west);
  \draw[arr] (leg) -- node[lbl, right] {valid key} (gw);
  \draw[arr] (gw) -- (api);
  \draw[arr] (oc) -- node[lbl, right, align=left] {webhook,\\verified\\source} (ad);
  \draw[arr] (ad) -- node[lbl, above, font=\tiny] {HMAC} (ctl);
  \draw[arr] (ctl) -- node[lbl, right, align=left] {revoke peer,\\15 min} (gw);
  \draw[arr] (ad) -- (col);

  \begin{scope}[on background layer]
    \node[draw, dashed, rounded corners=4pt, fit=(atk)(leg)(oc)(gw)(api)(ad)(ctl)(col), inner sep=7pt, label={[font=\scriptsize]below:internal Docker network, no published ports}] {};
  \end{scope}
\end{tikzpicture}
\caption{V2 architecture and trust boundaries. Enforcement happens in the gateway; the collector only stores redacted evidence.}
\label{fig:arch}
\end{figure}

\textbf{V2 defense.} A separate gateway (Fig.~\ref{fig:arch}) replaces the leak with controls under three conditions: \emph{baseline} (leak present, no deception), \emph{hardened} (debug route returns 404) and \emph{defended} (debug route returns a planted canary key, with canary enforcement). Canary use revokes the triggering peer for 15 minutes. A real OpenCanary login alert is accepted by an adapter only from the verified sensor address, redacted, signed with a per-run HMAC secret and posted to a gateway control endpoint, which rejects stale (older than 30\,s), forged and replayed events. Revocation applies to that peer's application requests even when it holds a valid key; other clients keep access.

\textbf{Threat model.} The attacker is an LLM agent with a path-only HTTP tool pinned to one lab origin, plus a submit tool. It may read any response and use any credential it finds. It cannot run code on the target, reach the internet or alter the sensor. The defender controls the gateway, the sensor and the adapter. Client identity is the socket peer (or, in Study~4, an HMAC-signed lab peer header); \code{X-Forwarded-For} is ignored.

\textbf{Harness and evidence.} Agents run under hard budgets on HTTP calls, model turns, reported tokens and wall time. Tokens come from the model runtime's usage reports and are never estimated. Every agent and defense event is appended to a SHA-256 hash chain; a verifier recomputes digests, coverage, summary counts and statistics. Logs are unsigned, so they detect inconsistency but do not attest authorship. Containers run on an internal Docker network with no published ports, as a non-root user with a read-only filesystem, no capabilities and resource limits. The local container lab passed 60/60 isolation and enforcement checks.

\section{Study 1: A Live Agent Bypasses Optional Decoys}
In both deception modes, the agent ignored every decoy and reached the genuine objective in 6 HTTP calls; no containment fired.

\textbf{Setup.} Model \code{tantular-cyber:latest} (9.7B parameters, Q8\_0, served by local Ollama), seed 13112026, one session per V1 mode, each against a fresh loopback instance of the Records API. Budgets: 25 HTTP calls, 30 model turns, 60{,}000 reported tokens and 180\,s wall time.

\begin{table}[t]
\centering
\caption{Study 1 outcomes (one session per mode).}
\label{tab:s1}
\small
\setlength{\tabcolsep}{3pt}
\begin{tabularx}{\columnwidth}{@{}lYrrrr@{}}
\toprule
Mode & Outcome & Calls & Tokens & Time & Decoys \\
\midrule
None & Wall budget reached & 18 & 32{,}414 & 182.3\,s & 0 \\
Static & Genuine objective & 6 & 8{,}030 & 64.6\,s & 0 \\
Adaptive & Genuine objective & 6 & 8{,}030 & 58.1\,s & 0 \\
\bottomrule
\end{tabularx}
\end{table}

In the static and adaptive sessions (Table~\ref{tab:s1}) the model read the genuine key at \code{/api/v1/debug/env} and called \code{/api/v2/records/export}; it never requested a decoy path. In the baseline session it also found the key but kept trying incorrect \code{/api/v1/...} export paths until the wall budget expired. All three event chains verified.

\textbf{Interpretation.} With $n=1$ per mode this is not a comparative result. It is a concrete counter-example to an optimistic answer to RQ1: when a genuine weakness is reachable, decoys placed beside it can simply be skipped. The baseline failure is a capability miss, not a defense effect. This observation motivated V2, which moves the bait onto the attacker's path by replacing the leak itself.

\section{Study 2: Controlled Model Study with an Ablation}
Hardening alone and hardening plus deception both reduced genuine exports from 20/20 to 0/20; deception added revocation of the attacking client in 19 of 20 sessions but no additional prevention.

\textbf{Protocol.} The protocol and source fingerprints were written and frozen locally before the first measured call (not externally preregistered). Model \code{qwen3.5:4b-q8_0} (local Ollama), temperature 0.2, 8{,}192-token context, 256 generated tokens per turn, thinking disabled. Three conditions (baseline, hardened, defended), 20 trials each, 60 sessions; condition order rotates across trials, and the three sessions of a trial share a sampling seed (from 131200) and the same task. Every session uses a fresh target and a new conversation, with random synthetic credentials. Budgets per session: 8 HTTP calls, 8 model turns, 30{,}000 reported tokens and 90\,s wall time. The task tells the agent the discovery routes and the export route, so the study measures exploitation of a known weakness rather than route discovery. Failures and timeouts are retained; none were replaced.

\textbf{Hypotheses and gate.} The primary outcome is paired genuine-objective success, baseline versus defended, tested with an exact two-sided McNemar test~\cite{mcnemar1947}. The prespecified gate requires at least 20 valid pairs, baseline success of at least 80\%, fewer successes under defense, $p<0.05$, and at most 10\% invalid runs per condition. Hardened versus defended is a secondary ablation, reported without claiming incremental benefit unless independently supported. Intervals are Wilson 95\%~\cite{wilson1927}.

\begin{table*}[t]
\centering
\caption{Study 2 results. All 60 sessions were valid.}
\label{tab:s2}
\begin{tabular}{@{}llccc@{}}
\toprule
Condition & Debug route & Genuine export success & Wilson 95\% CI & Peer revoked \\
\midrule
Baseline & Leaks the real lab key & 20/20 (100\%) & 83.9\%--100\% & 0/20 \\
Hardened & Returns 404 & 0/20 (0\%) & 0\%--16.1\% & 0/20 \\
Defended & Serves a planted canary key, enforcement on & 0/20 (0\%) & 0\%--16.1\% & 19/20 \\
\bottomrule
\end{tabular}
\end{table*}

\textbf{Primary test.} Baseline versus defended (Table~\ref{tab:s2}): 20 baseline-only successes and no reverse discordances; exact McNemar $p = 1.9073486328125\times10^{-6}$. The gate passes.

\textbf{Ablation.} Hardened versus defended: no discordant pairs; $p=1$. The study therefore supports the protected configuration as a whole but does not isolate any preventive benefit of deception for this model and task.

\textbf{The one unrevoked defended session.} In trial 11 (seed 131210) the model read the bait route, presented an unrecognised credential to the export (HTTP 401), then queried an absent debug route until its turn budget ended. It neither exported nor triggered revocation.

\textbf{Verification.} An independent verifier recomputed result and protocol digests, source fingerprints, session coverage, summary counts, Wilson intervals and the exact paired statistics across 60 sessions and 120 event chains. The study measured an archived gateway snapshot; later gateway edits (expiry cleanup, renewal on fresh alerts, weak-key rejection, 400/405 handling) were not part of it.

\section{Study 3: Cloud Validation of Alert-Driven Revocation}
On Google Cloud, the final gateway passed 14/14 control checks: a real OpenCanary login alert revoked the attacking client, whose valid token was refused 303\,ms (observed upper bound) after the login began, while a legitimate client completed 100/100 requests.

\textbf{Setup.} One \code{e2-small} VM (Ubuntu 24.04, \code{asia-southeast2-a}) with no service account, reachable only through Identity-Aware Proxy SSH, ran a separate V2 stack on an internal Docker network with no published ports. A runner generated fresh synthetic keys, executed the checks in a fixed order and removed the stack. The verifier deliberately handed the attacker container a valid lab token, so that the test measures whether detection, rather than the token, decides access.

\begin{table}[t]
\centering
\caption{Control checks shown in the final run's summary (14/14 passed in total).}
\label{tab:s3}
\small
\begin{tabularx}{\columnwidth}{@{}Yl@{}}
\toprule
Check & Observed \\
\midrule
Export without a credential & 401 \\
Legitimate client before detection & 200 \\
Attacker with supplied valid token, before detection & 200 \\
Forged sensor webhook from the attacker & 403 \\
Replay of the V1 debug-key bypass (separate client) & 403, revoked \\
Real OpenCanary login from the attacker & 200, alert \\
Same attacker, same valid token, after the alert & 403 \\
Spoofed forwarded identity after revocation & 403 \\
Legitimate client after revocation & 100/100 OK \\
Invalid HMAC signature & 401 \\
Signed control event & 202 \\
Replayed signed event & 409 \\
\bottomrule
\end{tabularx}
\end{table}

\textbf{Event chain.} OpenCanary logged the login from \code{172.19.0.2} at 08:14:17.957 UTC; the gateway logged \code{defense.revoked} for that peer at 08:14:17.970 and \code{defense.denied} on the export at 08:14:18.046. The sensor event and the revocation share one event ID, and the submitted password does not appear in saved telemetry.

\textbf{Timing.} The 303\,ms figure is measured from login initiation to the first observed denied export and includes request and polling overhead. An earlier gateway revision passed the same 14 checks with an observed bound of 266\,ms. These are two single observations, not a latency distribution.

\textbf{Regression.} The final gateway passed 9/9 focused tests, the full lab suite (79/79) and a strict TypeScript check. The VM and its disk were deleted after the run, and the deletion was confirmed by inventory.

\section{Study 4: Multi-Agent Swarm Engagements}
When agents used the canary, revocation was immediate and scoped to the user of the canary; across engagements, budget limits and agent capability decided as many outcomes as the defense did.

\textbf{Setup.} A swarm of role-prompted agents (recon, harvest, exploit, bait) and one legitimate agent attacked the V2 gateway on a GCP VM through an IAP tunnel. Each agent carried an HMAC-signed peer identity, so revoking one agent could not affect its siblings. A monitoring hub mirrored defense events; the gateway's chained log remained the record of truth. Models ran on the operator's machine via local Ollama.

\begin{table*}[t]
\centering
\caption{Swarm engagements on 14 November 2026 (one engagement each). Engagement~A used the deterministic scripted driver.}
\label{tab:s4}
\begin{tabularx}{\textwidth}{@{}llllcY@{}}
\toprule
Engagement & Agents & Budget & Red agents revoked & Real exports & Legitimate agent \\
\midrule
A: scripted, no model & 8 & default & 4 (harvest $\times$2, exploit $\times$2) & 0 & Exported normally before and after revocations \\
B: \code{qwen3.5:4b-q8_0} & 6 & default & 5 of 6 agents revoked in total & 0 & \textbf{Took the canary bait and was revoked} \\
C: \code{tantular-cyber} & 6 & 180\,s wall & 0 of 5 (all budget-stopped) & 0 & Budget-stopped; never exported \\
D: \code{tantular-cyber} & 6 & 40 calls, 50 turns & 3 of 5 & 0 & Saw the bait, did not use it; not revoked; never exported \\
\bottomrule
\end{tabularx}
\end{table*}

\textbf{Capture latency (D).} The three captured agents in Table~\ref{tab:s4} were revoked on first canary use, 39\,s, 54\,s and 59\,s after the bait was served. Their 58 later requests were denied, including retries of the canary minutes later, while other agents kept access.

\textbf{Confounds.} In C, six agents shared one local runtime and averaged about five model turns in 180\,s; containment there came from budgets, not the defense. In D, all 90 export requests targeted a non-existent \code{/api/v1/...} path, so zero exports reflect a capability miss as much as the defense. One agent lost 19 of 40 calls to the harness's own path guard.

\textbf{False-positive mode (B).} The legitimate agent held the correct key in its prompt, yet fetched the debug route, preferred the ``leaked'' credential, used it and was revoked. For deception aimed at LLM clients, bait that looks useful can also be taken by benign automation.

\textbf{Hub hardening found during the study.} Two issues in the monitoring hub (not the gateway) were found and fixed: a protocol-relative path could make its reverse proxy reach other internal hosts, and Docker bridge settings did not reliably block its egress. Public DNS resolution from the hub remained possible on GCP and requires a host-level rule. Neither affected the gateway's enforcement.

\section{Discussion}
Across four studies, deception's measurable contribution against LLM agents was fast, attributable detection; prevention came from closing the real weakness.

\textbf{RQ1: traps beside the path are optional.} The Study~1 agent never requested a decoy because a genuine leak answered its need first. Deception literature often assumes the intruder explores; a goal-directed agent with a working path does not need to. Bait works when it occupies the place the attacker must look, as in V2, where the canary replaced the leak and 19 of 20 defended sessions used it.

\textbf{RQ2: hardening did the preventive work.} The ablation is the central result: hardened and defended conditions were indistinguishable on exfiltration (0/20 each). A defender who removes the leak gets the prevention; adding deception adds a signal. This does not mean deception is useless. It identified and cut off the attacking client in 19 sessions, which the hardened condition, by design, could not do. That distinction matters for attribution, response and threat intelligence, and it is invisible in studies without a hardened arm.

\textbf{RQ3: alerts can drive narrow enforcement.} Study~3 shows that a commodity honeypot alert can be authenticated, redacted, signed and turned into per-client revocation within a few hundred milliseconds, without denying a co-resident legitimate client. Studies~3 and~4 also show what makes this safe in practice: a verified sensor source, a short validity window, replay protection and an identity the client cannot spoof.

\textbf{LLM clients change the false-positive calculus.} The premise that only attackers touch decoys assumes benign clients follow instructions. Engagement~B contradicts it: a legitimate LLM agent preferred a leaked-looking credential to its assigned one. Deployments that serve AI-driven integrations should treat bait use as high-signal but not conclusive, and favour reversible, scoped responses such as the 15-minute peer revocation used here.

\textbf{Capability confounds outcomes.} In several runs, small local models failed for reasons unrelated to the defense (wrong API version, budget exhaustion, harness guard hits). Zero exports must therefore always be read alongside evidence of \emph{why}; we recorded explicit \code{canary.used}, \code{defense.revoked} and denial events rather than inferring containment from a null answer.

\section{Limitations, Ethics and Artifacts}
\label{sec:limits}
Every result is bounded by one synthetic target, small samples and two small local models; none is a claim about production systems.

\textbf{Threats to validity.}
\begin{itemize}
\item \emph{External.} One mock API with one planted weakness; local models of 4B and 9.7B parameters; one runtime and host per study. Results may not transfer to larger or commercial models, other vulnerability classes or real applications.
\item \emph{Construct.} Study~2 tells the agent the discovery and export routes, so it measures exploitation of a known weakness, not discovery. Its task differs from Study~1, so the two are not numerically comparable.
\item \emph{Statistical.} Study~2 has 20 pairs; Wilson intervals for 0/20 still allow up to 16.1\% success. Studies 1, 3 and 4 are single runs or single engagements and are exploratory.
\item \emph{Internal.} Study~2 measured an archived gateway snapshot; Study~3 measured a later revision. Budgets and shared inference shaped several swarm outcomes.
\item \emph{Pre-registration.} Study~2's protocol was frozen locally before measurement but not externally timestamped or reviewed.
\item \emph{Evidence integrity.} Hash chains and digests detect inconsistency but are unsigned; a party able to replace all evidence could recompute them.
\end{itemize}

\textbf{Production gap.} IP- or header-based peer identity can affect clients behind shared NAT; revocation and replay state are in memory; events are not shipped to durable audit storage; there was no load testing, representative benign traffic or independent adversarial review.

\textbf{Ethics and safety.} All targets, records and credentials were synthetic, and no third-party system was attacked. Containers ran on internal networks with no published ports; agents were limited to path-only requests against one pinned origin, with hard budgets and egress guards. Commercial model APIs were not used. Decoys contained no prompt-injection or hidden instructions. The cloud VM had no service account, was reachable only via authenticated IAP SSH, stopped itself after a fixed runtime and was deleted after testing.

\textbf{Artifact availability.} The lab source, protocol, result files with chained logs, verifier and evidence transcripts are publicly available at \url{https://github.com/ghifiardi/AI-Deception-Lab}; the exact version used here (release v1.0-preprint) is archived at \url{https://doi.org/10.5281/zenodo.24217354}. The Study~2 result can be re-verified with the bundled verifier and reproduced with local model weights.

\section{Conclusion}
Against LLM agents, a trap only works if it stands where the agent must go. In our lab a live model walked past optional decoys to a genuine leak; once the leak was closed and baited, exfiltration stopped, but it stopped equally well with hardening alone. What deception added was a high-confidence, attributable signal that a real honeypot alert could turn into narrowly scoped revocation in a few hundred milliseconds, without disrupting a legitimate client. We also saw a benign LLM client take the bait, a reminder that ``only attackers touch decoys'' weakens when clients are themselves agents. Future work should test larger and commercial models, multiple targets and weakness classes, realistic benign AI traffic, and identity bound to authenticated sessions, with a hardened arm in every comparison.

\end{document}